\documentclass[fleqn,10pt]{wlscirep}
\usepackage[utf8]{inputenc}
\usepackage[T1]{fontenc}

\newcommand{\beq}{\begin{equation}}
\newcommand{\eeq}{\end{equation}}
\newcommand{\bea}{\begin{eqnarray}}
\newcommand{\eea}{\end{eqnarray}}
\newcommand{\be}{\begin{equation}}
\newcommand{\ee}{\end{equation}}

\title{Multiwave pandemic dynamics explained: \\ 
How to tame the next wave of infectious diseases}

\author[1,2,*,+]{Giacomo Cacciapaglia}
\author[1,2,+]{Corentin Cot}
\author[3,4,*,+]{Francesco Sannino}
\affil[1]{Institut de Physique des 2 Infinis (IP2I), CNRS/IN2P3, UMR5822, 69622 Villeurbanne, France}
\affil[2]{Universit\' e de Lyon, Universit\' e Claude Bernard Lyon 1, 69001 Lyon, France}
\affil[3]{CP3-Origins \& the Danish Institute for Advanced Study, University of Southern Denmark, Campusvej 55, DK-5230 Odense, Denmark}
\affil[4]{Dipartimento di Fisica E. Pancini, Universit\`a di Napoli Federico II \& INFN sezione di Napoli, Complesso Universitario di Monte S. Angelo Edificio 6, via Cintia, 80126 Napoli, Italy}

\affil[*]{g.cacciapaglia@ipnl.in2p3.fr, sannino@cp3.sdu.dk}

\affil[+]{these authors contributed equally to this work}

\begin{abstract}
\end{abstract}
\begin{document}

\flushbottom
\maketitle
\thispagestyle{empty}

{\bf 
Pandemics, like the 1918 Spanish Influenza~\cite{1918influenza} and COVID-19, spread through regions of the World in subsequent waves. 
There is, however, no consensus on the origin of this pattern, which may originate from human behaviour rather than from the virus diffusion itself. Time-honoured models of the SIR type~\cite{Kermack:1927} or others based on complex networks~\cite{WANG20151,PERC20171,ZHAN2018437} describe well the exponential spread of the disease, but cannot naturally accommodate the wave pattern. 
Nevertheless, understanding this time-structure is of paramount importance in designing effective prevention measures.
Here we propose a consistent picture of the wave pattern based on the epidemic Renormalisation Group (eRG) framework~\cite{DellaMorte:2020wlc,Cacciapaglia:2020mjf}, which is guided by the global symmetries of the system under time rescaling.
We show that the rate of spreading of the disease can be interpreted as a time-dilation symmetry, while the final stage of an epidemic episode corresponds to reaching a time scale-invariant state. We find that the endemic period between two waves is a sign of instability in the system, associated to near-breaking of the time scale-invariance. This phenomenon can be described in terms of an eRG model featuring complex fixed points~\cite{cacciapaglia2020evidence}. 
Our results demonstrate that the key to control the arrival of the next wave of a pandemic is in the strolling period in between waves, i.e. when the number of infections grow linearly. Thus, limiting the virus diffusion in this period is the most effective way to prevent or delay the arrival of the next wave.
In this work we establish a new guiding principle for the formulation of mid-term governmental strategies to curb pandemics and avoid recurrent waves of infections, deleterious in terms of human life loss and economic damage. }

\vspace{1cm}

As it emerged from the Spanish Influenza that hit the World in three consecutive waves between spring 1918 and the early months of 1919, virus-driven pandemics can feature a wave pattern, even though the origin of this behaviour is not understood \cite{1918influenza}. The very recent pandemic, caused by the coronavirus SARS-CoV-2, is showing a similar pattern, with a first wave hitting in the spring of 2020, and following ones still raging various regions of the World. Reliable algorithms were used at the beginning of the pandemic to predict the evolution of the number of cases affected by the COVID-19 disease \cite{Perc2020,Hancean2020,Zhou2020}, however it has proven difficult to predict the arrival of a second wave in the fall 2020 \cite{Scudellari}. With the exception of a few countries like China, Vietnam and New Zealand, all regions of the World are suffering from multiple waves of COVID-19 infections.

The diffusion of the virus can be described by various time-honoured models, like compartmental models of the SIR type \cite{Kermack:1927} and complex networks \cite{WANG20151,PERC20171,ZHAN2018437}. These mathematical frameworks account for the exponential growth of the number of new infected cases, and the slowing down of the spreading once most of the susceptible cases are infected. However, it is not a simple task to generate a wave pattern. For instance, in SIR models, one could induce a second wave either by injecting by hand new individuals in the susceptible sub-population, or by including a probability that the removed individuals may return to the state of susceptible. The latter case cannot apply to the COVID-19, as very few cases of recovered individuals being infected again have been recorded.

In an article first posted at the beginning of August \cite{cacciapaglia2020second}, we successfully predicted the occurrence of a second wave in Europe starting in September--October. The analysis is based on the eRG framework \cite{DellaMorte:2020wlc}, extended by interactions between various countries \cite{Cacciapaglia:2020mjf}. The approach is based on the analysis of the time evolution of the total number of infected cases and the symmetries that this epidemic curve reveals, allowing to extract reliable  information from the data independently on the specific conditions met in each country. In fact, all the elements that can influence the velocity of the disease spreading are included in a single parameter, which contains the effect of local conditions, non-pharmaceutical interventions and socio-demographical characteristics. 
The eRG, therefore, can provide complementary information to studies that analyse in detail the effect of various measures \cite{Lai2020,Flaxman2020,Chinazzi,SEIR,scala2020}.
As an example, the eRG framework has been used to study the effect of mobility reduction in Europe and the US during the first wave \cite{cacciapaglia2020mining}, highlighting a universal time-frame of 2-4 weeks before an observable effect can be detected in the virus diffusion. For comparison, detailed studies of the mobility in the US \cite{Leskovec2020} have been able to identify the locations and events that foster the infection of new individuals and ignite hotspots.

In this work we focus on the total number of infected cases, as this is the most reliable tracker of the time-evolution of the pandemics. In fact, other data, like the number of deaths and of hospitalisations, depend on factors like the age distribution and medical pre-conditions of the infected individuals, which can influence the delay between the infection and the time-stamp in the data.
The master multiwave equation for the time-evolution of the total number of infected cases $I_j (t)$ in a region $j$ reads:
 \begin{equation}  \label{eq:multiwave}
- \beta_{\rm multiwaves} (I_j) = \frac{1}{A_j} \frac{d I_j(t)}{\gamma_j\ d t} =   \frac{I_j}{A_j} \left[  \left(1 - \frac{I_j}{A_j} \right)^2  - \delta_{j,0} \right]^{p_{j,0}} \; \prod_{\rho=1}^w \left[\left( 1-\zeta_{j,\rho} \frac{I_j}{A_j} \right)^2 - \delta_{j,\rho} \right]^{p_{j,\rho}} +\  \sum_l \frac{k_{jl}}{\gamma_j\ n_{m j}}  \frac{ {I}_l - {I}_j}{A_j}\ ,
\end{equation}
where the first term on the right-hand side is a generalisation for $w+1$ consecutive waves of the CeRG equations \cite{cacciapaglia2020evidence} and the second term contains the interactions between regions \cite{Cacciapaglia:2020mjf}. Here, we will always consider the number of cases per million inhabitants in order to compare different regions. In the master equation, most of the parameters are explicitly independent on the normalisation, as $I_j(t)$ always appears divided by the total number of cases at the end of the first wave, $A_j$: the only exception is the interaction term, which also depends on the population of the regions ($n_{m j}$ measures the population of region-$j$ in millions).
The parameters $\gamma_j$ measure the effective velocity of the virus in each regions, and can be associated to an effective infection rate. This parameter can be eliminated from the equation by measuring the time in terms of a region-dependent scale, $\tau_j = \gamma_j t$, once the couplings $k_{jl}$ are also rescaled: $\gamma_j$ can therefore be interpreted as a local \emph{time-dilation}, characteristic of each region and taking into account all the non-pharmaceutical measures and local conditions in each region. 
These parameters can be extracted from the data at the beginning of the epidemic diffusion in each region, independently on the normalisation of the number of cases, which is very sensitive to the testing strategies \cite{testing} changing during the pandemic. More details on the equations, and on the meaning of other parameters can be found in the supplementary material.

The master equation \eqref{eq:multiwave} encodes the multi-wave pattern in two ways: in the first term, the parameters $\delta_{j,\rho}$ destabilise the fixed points at $I_j (\tau^\ast_\rho) = A_j/\zeta_\rho$; in the second term, the interactions with other regions, or with an external source, can also destabilise the system and drive a new growth of $I_j$ towards the next fixed point. In fact, for $\delta_{j,\rho} = 0$, the number of infected will grow until $I_j (\tau_\rho^\ast) = A_j/\zeta_\rho$, where the growth stops because of the vanishing of the beta function. This is a steady-state, independent of time, which signals the end of the infection. For $\delta_{j,\rho} < 0$, the zero is moved on the complex plane and cannot be reached, thus driving an endemic state with linear growth, which we call \emph{strolling} in honour to the application of this formalism in high energy physics \cite{cacciapaglia2020evidence}.
The second term has been used \cite{cacciapaglia2020second} to predict the European second wave of September 2020, where $k_{jl}$ was associated to an estimated number of travellers between each country. In general, both effects are expected to be present: as we will see, the instability due to $\delta$ provides a maximal delay for the arrival of the next wave, which is directly related to the number of new cases recorded during the strolling period between waves. The presence of a large interaction can induce an early arrival of the new wave.

Thus, effective measures to prevent and control the next wave of a pandemic like COVID-19 can only go via a strict control of the number of cases inside the country or region, combined to effective tracking of new infected individuals traveling in. This is precisely the strategy followed by China, Vietnam and New Zealand, leading to an early extinction of the disease and the absence of a second wave.

\section*{Results}

\begin{figure}[tb!]
\begin{center}
\includegraphics[width=.99 \textwidth]{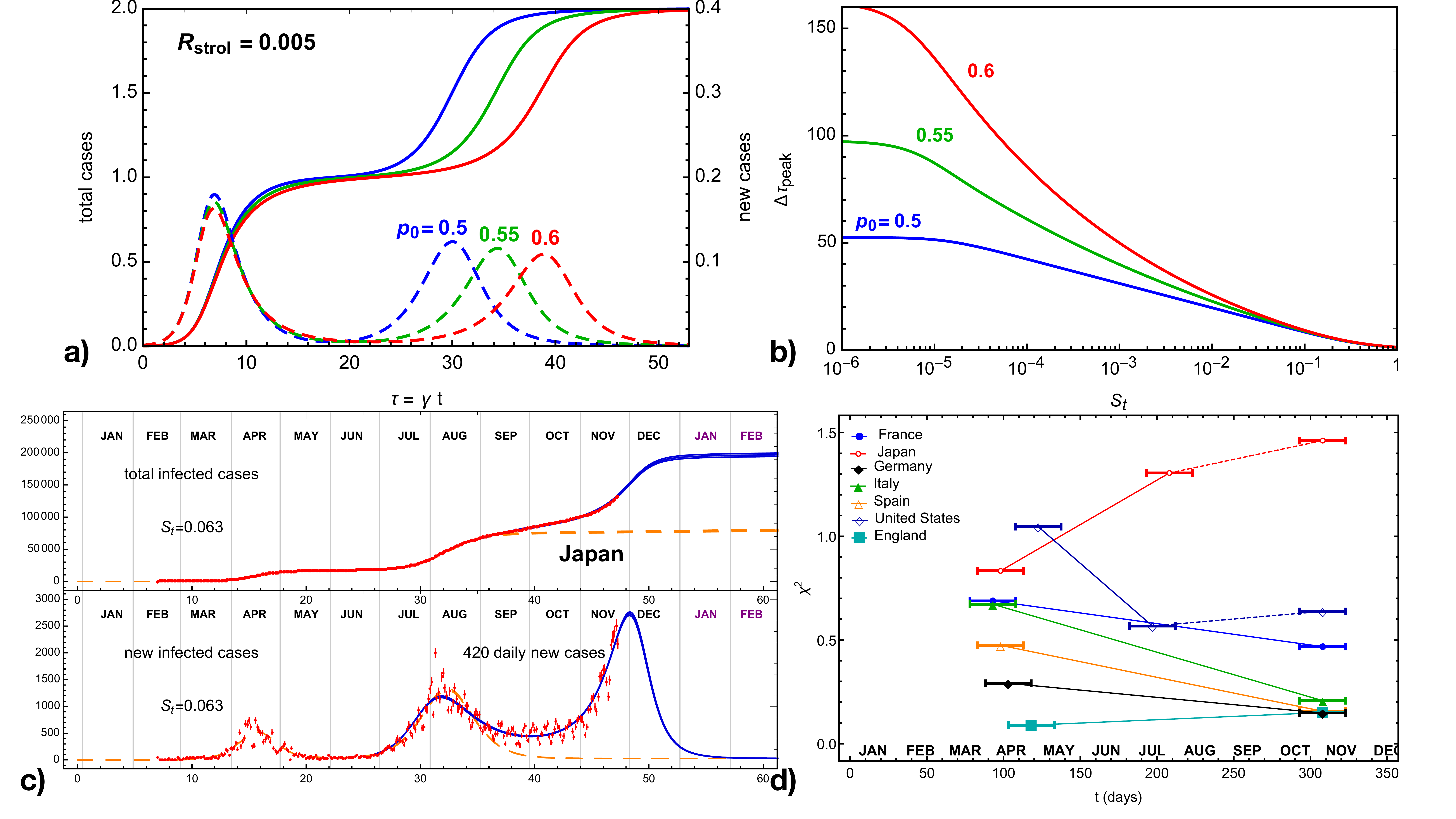}
\end{center}
\caption{{\bf Illustration and validation of the CeRG multiwave model.} The panel a) shows sample solutions of the CeRG multiwave equation for $w=1$, with $\delta_1 = 0$, so that the epidemic episode is extinguished after two waves. The total number of cases, normalised to the first peak, is shown in solid, while the normalised new cases are shown in dashed. In the panel b) we show the dependence of the delay between the two peaks of new infections, $\Delta \tau_{\rm peak}$ measured in the local time, as a function of $S_{\rm t}$. The CeRG parameters are fixed to the following values, unless specified: $p_0 = 0.5,\ p_1 = 0.65,\ \zeta_1 = 0.5$. Panel c) shows the CeRG model applied to the second and third wave in Japan (blue) as compared to the data (red) and the eRG fits of the two waves (see Table \ref{tab:1}).  In panel d) we show the value of the geographical uniformity indicator as defined in the text for a sample of countries, showing that the virus is more equally spread in the various regions during the second wave, in most cases.}
\label{fig:1}
\end{figure}

The CeRG multiwave model, corresponding to~\eqref{eq:multiwave} without the interaction terms, can be used to effectively describe a pandemic episode in multiple subsequent waves. To illustrate the model, in the top panels of Fig.~\ref{fig:1} we show some features of the solutions for an isolated region with only two waves, corresponding to $w=1$ and $\delta_1 = 0$. The latter condition ensures that the epidemic is extinguished after the second wave. All results are shown for number of cases normalised to the first wave, and expressed in the local time $\tau = \gamma t$. Panel a) shows three solutions (solid lines) together with the corresponding new cases (dashed lines), for three values of $p_0$. We can clearly see the two-wave structure emerging in the solution, the fact that the second peak tends to be flatter than the first (for $\zeta_1 = 1/2$). Furthermore, larger values of the exponent $p_0$ tend to delay the second peak and flatten it.
An important factor in controlling the arrival of a future wave is the number of new cases during the intermediate strolling phase, which we encode in the parameter
\beq \label{eq:Rstrol}
S_{\rm t} = \left. \frac{1}{A \gamma} \frac{d\ I}{d\ t} \right|_{\rm strol}\,,
\eeq
 normalised to the total number of infected cases after the first wave, $A$, and expressed in terms of the local time. This is a crucial parameter in controlling the timing of the second wave, as illustrated in panel b) of Fig.~\ref{fig:1} where we show the peak delay $\Delta \tau_{\rm peak}$ as a function of $S_{\rm t}$. For $S_{\rm t} \to 1$, the delay goes to zero as the two waves merge into a single one, while it grows for smaller values following a power law that depends on $p_0$.
Remarkably, for small enough $S_{\rm t}$, the peak delay saturates to a finite value, indicating that the second wave cannot be pushed away as it seems inevitable as long as a strolling period is present. The peak delays also depend strongly on $\zeta_1$, which encodes the height of the second peak, being more delayed for decreasing $\zeta_1$: for values above $0.8$, however, the second wave becomes too small, thus the solution looses physical relevance.

Comparing the predictions of the CeRG multiwave model to data is not an easy task: in fact, the number of detected infections, collected via the positivity of the tests done in each country, depends crucially on the number of tests done each day \cite{testing} and on the specific testing policy adopted over time. 
For instance, during the first wave, many countries did fewer tests while focusing on hospitalised cases, while a more extensive testing campaign occurred starting in the summer months. As a consequence, quantities that depend on the number of cases, like the delay between peaks, cannot be computed accurately as a bias between two waves is present in the data. 
We, therefore, will apply the model to the second wave and following ones. As an example, in panel c) of Fig.~\ref{fig:1} we show the data for Japan compared to a scenario based on the CeRG multiwave model. Japan is an ideal candidate for this model since, being an island, the frontiers can be well-controlled, and one can consider the country as an isolated system. Furthermore, the second wave has already ended, followed by a two-month period of strolling with around 450 new infected cases detected each day. The CeRG model, shown in blue, provides a good quantitative and qualitative description of the data, and predicts that the third wave will peak at the beginning of December and be slightly higher than the second ($\zeta_1 = 0.4$). 
Between the first and second waves, instead, no significant strolling was observed. This scenario could be interpreted in the following way: after the first wave, the virus diffusion was strongly limited by the enacted measures. However, new infected cases may have entered the country from abroad and/or spontaneous emergence of local hotspots inside the country (parametrised by the $k$-interactions). After the second wave, the virus kept spreading geographically within the population triggering at a later stage the third wave. The latter phase can be described by the CeRG model, while the transition between the first and second wave is due to {\it external} interactions. We do not attempt to unify the three waves because of the bias in the counting.

Testing the geographical diffusion of the virus in each country can provide useful indications on the mechanism behind each wave. For this purpose, we define a uniformity indicator, $\chi^2$, which encodes how far is the distribution of new cases  in regions of the country from a completely flat one. Smaller values of $\chi^2$ indicate a more uniform distribution. For Japan, we considered the new cases in the various prefectures (where we exclude Okinawa for the geographical distance from the mainland) during the first and second waves, as shown in red in Fig.\ref{fig:1} d).  The second wave has a larger $\chi^2$, which could be interpreted as more localised diffusion due to hotspots or travellers returning to their home cities. We also report the indicator for the month of November, which we would expect to be reduced if the strolling plays an important role in creating the third wave. The result is too preliminary, as Japan is still far from the peak of the third wave and the indicator is found to be minimised at the peak. 

Having validated the CeRG approach, we can now apply it to understand and predict the next wave in various regions of the World. As a caveat, we should recall that the interactions between regions and the presence of hotspots can also affect the results and anticipate the insurgence of the next wave.

\begin{table}[tb!]
\begin{tabular}{|c|c|c|c|c|c||c|c|c|}
\hline
 & \multicolumn{5}{c||}{eRG fits} & \multicolumn{3}{c|}{CeRG forecast} \\ \hline
country & \multicolumn{2}{c|}{1$^{\rm st}$ wave} & strolling & \multicolumn{2}{c||}{2$^{\rm nd}$ wave} & 2$^{\rm nd}$ peak  &  & 3$^{\rm rd}$ peak\\ \hline
             & $A$ & $\gamma_{\rm eRG}$  & $I'_{\rm strol}$ & $A$ & $\gamma_{\rm eRG}$  &  date ($\pm 1$ w) & $I'_{\rm strol}$ &  date  ($\pm 1$ w)\\ \hline
France & $2034(13)$ & $0.135(3)$ & $5.2(8)$ & $61(6) \cdot 10^3$ & $0.048(5)$ & 2020-11-07 & $35$ & 2021-11-20 \\
Italy & $3638(28)$ & $0.100(3)$ & $3.5(4)$ & $41(2) \cdot 10^3$ & $0.075(8)$ & 2020-11-17 & $37$ & 2021-07-17 \\
UK & $3943(25)$ & $0.0794(15)$ & $9.1(2)$ &  $21.3(2) \cdot 10^3$ & $0.0717(8)$ & 2020-11-08 & $17$ & 2021-07-24 \\
Germany & $2006(14)$ & $0.128(3)$ & $5.1(2)$ & $20(3) \cdot 10^3$ & $0.058(2)$ & 2020-12-13 & $14$ & 2021-10-05\\
Spain  & $5000(30)$ & $0.134(4)$ & $7(1)$ &  $30(4) \cdot 10^3$ & $0.067(1)$ & 2020-10-31 & $22$ & 2021-08-04 \\
Switzerland & $3304(22)$ & $0.160(4)$ & $10(2)$ & $40(4) \cdot 10^3$ & $0.10(1)$ & 2020-11-04  & $70$ & 2021-04-20\\
Netherlands & $2546(15)$ & $0.112(2)$ & $10.3(2)$ & $26.1(2) \cdot 10^3$ & $0.0797(8)$ & 2020-10-24  &$24$ & 2021-06-19 \\
Belgium & $4758(27)$ & $0.116(2)$ & $22.1(6)$ & $40.7(4) \cdot 10^3$ & $0.121(2)$ & 2020-10-24 & $70$ & 2021-03-13 \\
Denmark  & $1950(10)$ & $0.088(2)$ & $6.7(1)$ &  $12(2) \cdot 10^3$ & $0.062(2)$ & 2020-11-07  &  $9$  & 2021-09-11   \\
Iceland & $5285(16)$ & $0.176(2)$ & $3.6(2)$ & $9.9(1) \cdot 10^3$ & $0.087(2)$ & 2020-10-10 & $12$ & 2021-04-20\\
Canada  & $2643(20)$ & $0.0732(14)$ & $8.2(2)$ & $16(2) \cdot 10^3$ & $0.046(5)$ & 2020-12-07 & $71$ & 2022-01-08 \\ 
\hline
South Africa & $11039(35)$ & $0.0704(5)$ & $26.69(9)$ & $-$ & $-$ & 2020-07-11 & $18$ & 2021-01-06 \\
Bolivia  & $12240(20)$ & $0.0442(2)$ & $9.2(3)$ & $-$ & $-$ & 2020-07-25 & $3.7$ & 2020-07-28 \\
Saudi Arabia  & $9560(20)$ & $0.0447(3)$ & $11.51(3)$ & $-$ & $-$ & 2020-06-20 & $9.2$  & 2021-04-17\\
\hline
Australia & $255(1)$ & $0.239(5)$ & $0.38(2)$ & $735(3) $ & $0.1015(7)$ & 2020-07-25 & $1$ & 2020-04-17 \\
Japan & $130.7(7)$ & $0.125(2)$ & $0.22(2)$ & $434(5)$ & $0.094(1)$ & 2020-08-08 & $3.4$ & 2020-12-02\\
South Korea & $185(2)$ & $0.225(14)$ & $0.569(8)$ & $157(3)$ & $0.136(6)$ & 2020-08-26 & $0.7$ & 2020-11-25\\ 
\hline
\end{tabular}
\caption{{\bf Numerical results for the eRG fit and the CeRG forecast for the next wave.}  Columns 2 to 6 contain the results for the fit of the first two waves using the eRG model \cite{DellaMorte:2020wlc}, where $A$ are indicated in number of cases per million inhabitants, $\gamma_{\rm eRG}$ is given in inverse days, and the new parameter $I'_{\rm strol}$ indicates the number of daily new cases per million inhabitants obtained by fitting the strolling period with a linear growth. 
The values highlight that the second wave typically features a smaller infection rate and a larger number of infected cases. In the last three columns we show the expected second and third peak dates (within a one week error) and the expected value of $I'_{\rm strol}$. Except for Australia, South Africa, Bolivia, Saudi Arabia, Japan and South Korea, we fix $S_{\rm t} = 0.01$, $p_0 = 0.55$, $p_1 = 0.6$ and $\zeta_1 = 1/2$, while $\gamma$ and $A_0$ are chosen to fit the second wave (first for South Africa, Bolivia, Saudi Arabia). For countries currently in the strolling period, all the parameters of the CeRG model are fitted.  \label{tab:1}}
\end{table}

\begin{figure}[tb!]
\begin{center}
\includegraphics[width=.99 \textwidth]{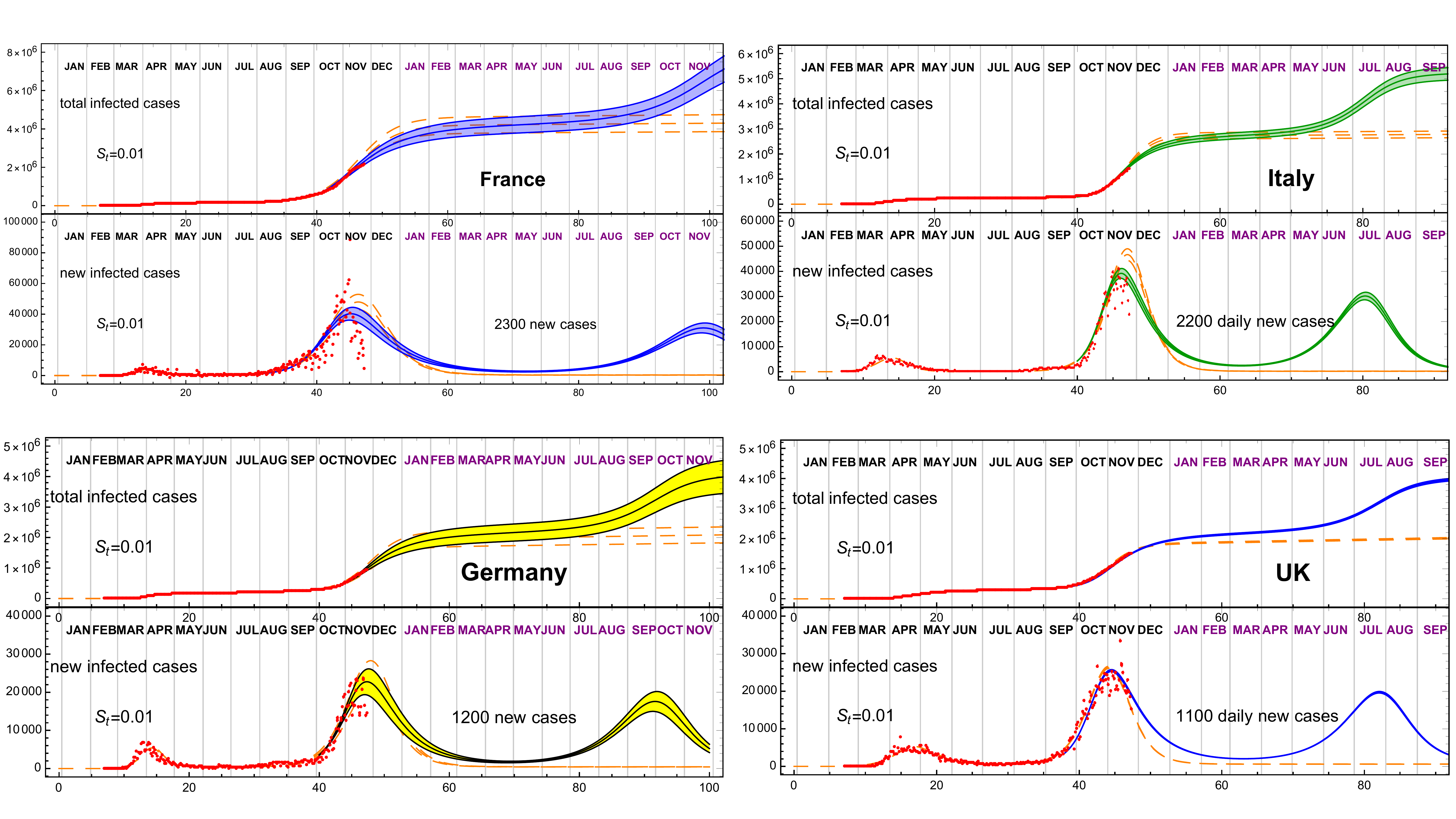}
\end{center}
\caption{{\bf Strolling as a precursor of a COVID-19 third wave.} Most European countries are still undergoing a second wave of COVID-19. In this figure we show how a strolling period, consistent on a fixed number of new infections per day (indicated in each panel) could lead to a third wave, as indicated by the solid band. The prediction corresponds to $S_{\rm t} = 0.01$, and is compared to the data (adjourned to November 23) and the eRG fit from Table \ref{tab:1} (dashed orange). The plot includes France, Italy, Germany and the UK.}
\label{fig:2}
\end{figure}

\begin{figure}[tb!]
\begin{center}
\includegraphics[width=.99 \textwidth]{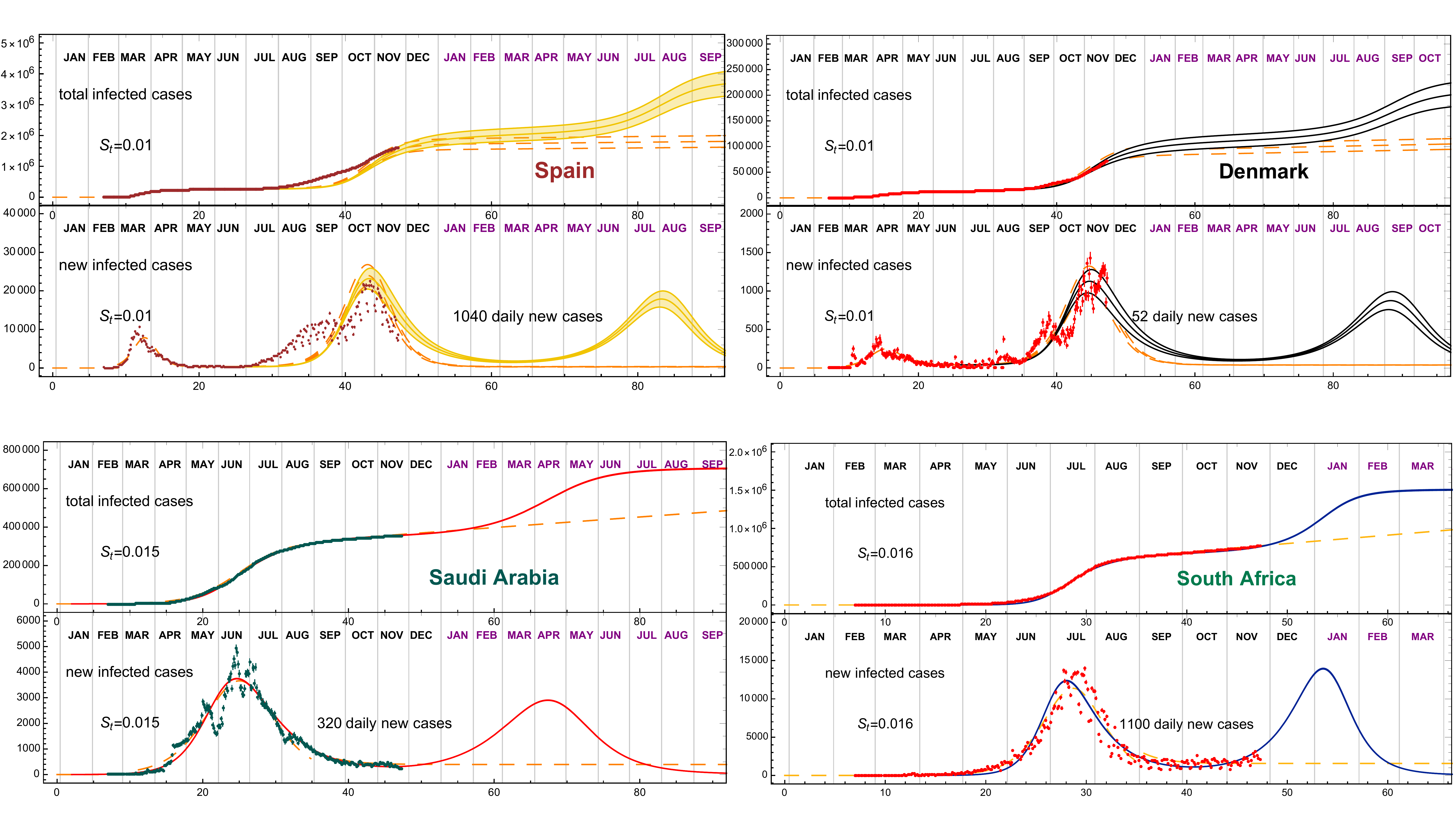}
\end{center}
\caption{{\bf Strolling as a precursor of a new COVID-19 wave.} The top panels shows Spain and Denmark, while the remaining European countries included in this study are shown in the supplementary material. In the bottom panels we show two sample countries from other regions of the World. In both cases, currently the epidemic is in the strolling regime after the first wave, indicating an imminent restart of the epidemic. In cases with an ongoing strolling, $S_t$ is fitted to reproduce the data.}
\label{fig:3}
\end{figure}

\paragraph{Europe:}
Most of the European countries are being hit by a second wave of the COVID-19. A general trend we observe is that the infection rate during the second wave is smaller than the one of the first wave, as shown by the values of $\gamma_{\rm eRG}$ in Table~\ref{tab:1}. The comparison is done by fitting the two waves independently by use of the eRG model. Moreover, the expected cumulative number of infected cases during the second wave is much larger than that for the first wave, even taking into account the higher number of tests performed during the second wave.
The emergence of the second wave was explained as arising from the interactions between countries \cite{cacciapaglia2020second}, nevertheless the presence of strolling between the two waves, shown in terms of the number of new cases per million inhabitants $I'_{\rm strol}$, indicates that both effects participated.  
The decrease of the geographical uniformity indicator, shown in Fig.~\ref{fig:1} d), indicates that the strolling had an important role in diffusing the virus across the countries (for England, the uniformity was present since the first wave). The geographical diffusion may also be the reason behind the fact that the second wave has infected a larger portion of the population. 

The CeRG model cannot be applied straightforwardly to the first two waves, due to the uncertainty in their relative normalisation, while it can be used to estimate when a third wave will hit. The result is shown in the last three columns of Table~\ref{tab:1}, and in Fig.~\ref{fig:2} and the top row of Fig.~\ref{fig:3} (additional plots are provided in the supplementary material). 
The timing of the third wave peak depends crucially on the time-dilation parameter $\gamma$, and on the amount of strolling in the intermediate endemic phase: in the projections, we fitted $\gamma$ to reproduce the second wave and fixed $S_{\rm t} = 0.01$, with the corresponding total number of new cases, expected during the strolling, reported in the figures. 
The last column of Table~\ref{tab:1} shows that a wide range of peak timings are expected, ranging from March to November 2021, where we associated an error of 1 week to the projection due to a variation of 10\% in the infection rate.
These results show clearly that controlling the infection rates and reducing the level of strolling after the end of the second wave are keys to delaying the next wave. Another element that should be included is the number of travellers across various countries \cite{cacciapaglia2020second}, which can help propagate the wave from country to country, thus affecting the ones with  pronounced delayed projections.

\paragraph{The US:}

The US has already seen two waves in April and July-August, and is undergoing a third. However, the first two waves are geographically distinct, with the episode in April mainly involving New York and New England, and the second spreading all over the remaining states. This is well illustrated by the geographical uniformity indicator in Fig.~\ref{fig:1} d), which sharply drops between the two episodes. The third point, based on the data of November, is still preliminary and will decrease as the third wave approaches its peak. To analyse the evolution of the COVID-19 epidemic in the US, a dedicated study which takes into account sub-regions is required.
The uniformity analysis suggests that the first two episodes should be described in terms of interactions between states, while the third one may be originated by the strolling. Results of this analysis and projections for the future waves in the US will be presented in a separated article \cite{futurepublication}.

\paragraph{Other countries:}

We included in our analysis a selection of countries for other regions of the World, selected in order to represent all continents. Note that we retained only countries for which the multi-wave analysis is best explained in terms of the CeRG model, i.e. where diffusion of the virus in sub-regions do not produce features that would require a multi-region analysis. The latter situation can be tackled within coupled CeRG equations, but this analysis goes beyond the scope of the present work.

In Table~\ref{tab:1} we show the results for the selected countries, also illustrated in the bottom row of Fig.~\ref{fig:3} for South Africa and Saudi Arabia.
In most selected cases, the country is in the strolling regime, following the end of the first or second wave, thus allowing us to tune the CeRG parameters to reproduce the strolling and give a more reliable forecast for the following wave. In some cases, like South Africa, the high level of strolling indicates that a new wave is imminent. As we are not trying to perform a fit, due to the many uncertainties in the social distancing and testing policies, the scenario we present should be considered as a probable one. Yet, it should be noted that  the model does not leave much room to modify the expected number of total infected cases during the future wave nor change the timing substantially, without drastic pharmaceutical or non-pharmaceutical interventions

\section*{Discussion}

We provide a mathematical understanding of the wave pattern for pandemics, like the COVID-19 one. The approach is employed to forecast the timing of a future wave based on the number of new infections during the intermediate endemic phase. The timing of the new wave is related to a newly introduced parameter, $S_t$, that can be easily deduced from the cumulative number of infected cases. 
We studied several countries in different regions of the World and, in absence of any pharmaceutical interventions, we  estimate the timing of the next wave of infections. We found countries where a new wave will start in December 2020, like South Africa, and countries where it could start as late as October 2021, like in France. Our predictions will be affected by the border control regulations with the generic effect of inducing an early increase in the number of infections.

Our understanding of the wave structure of the COVID-19 pandemic draws the attention to the inter-wave \emph{strolling} period. We discover that controlling the number of new infections during the strolling period is necessary to delay the beginning of a future wave. This amounts to imposing social distancing measures and break potential chains of infections after the end of the wave in order to keep $S_t$ as low as possible. Delaying the next wave is crucial in order to have enough time to realise an effective vaccine campaign. 

Our results can effectively guide policymakers to time (non)pharmaceutical interventions to delay or reduce the impact of future COVID-19 waves. Until now, most measures are taken when the number
of new infected cases is already grown substantially. At this point in time one can only contain the wave not avoid it with serious impact for the loss of human lives as well as the economy.  We prove that
intervening during the strolling period of endemic diffusion is essential to delay or avoid a new wave while buying time 
for pharmaceutical interventions, like an effective vaccine campaign. More specifically, to maximise the delay, the strolling
parameter must be kept small, $S_t \approx 10^{-5}$ for an optimal use of the enacted measures.  In most countries, this implies
that the number of new cases at the end of the wave should be kept at the level of a few units per day. This
effect can be achieved by keeping or introducing new measures after the end of the wave, in the form
that is more appropriate for the local conditions.

\section*{Acknowledgements}

G.C. and C.C. acknowledge partial support from the Labex-LIO (Lyon Institute of Origins) under grant ANR-10-LABX-66 (Agence Nationale pour la Recherche), and FRAMA (FR3127, F\'ed\'eration de Recherche ``Andr\'e Marie Amp\`ere'').

\section*{Author contributions statement}

This work has been designed and performed conjointly and equally by the authors. G.C., C.C. and F.S. have equally contributed to the writing of the article.

\section*{Additional information}

The authors declare no competing interests.
All data used in this work are obtained from open-source repositories: \href{https://ourworldindata.org/}{Ourworldindata.org},  \href{https://worldometer.info/}{Worldometer.info}, \href{https://www.citypopulation.de}{Citypopulation.de}, \href{ https://www.data.gouv.fr/fr/datasets/donnees-hospitalieres-relatives-a-lepidemie-de-covid-19/}{Data.gouv.fr}, \href{https://toyokeizai.net/sp/visual/tko/covid19/en.html}{Toyokeizai.net}.

\newpage

\section*{Methods}

\subsection*{CeRG approach to multi-wave dynamics}

 In the original \emph{eRG} approach \cite{DellaMorte:2020wlc}, rather than the number of cases, it was used its natural logarithm  
$\alpha(t) = \ln {I}(t)$. For a single wave pandemic this provides a good fit to the data. 
To describe multiple waves \cite{cacciapaglia2020evidence}, it is better to use the eRG directly for the cumulative number of total cases rather than its log. 
The derivative of $I(t)$ with respect to time is interpreted as the {\it beta function} of an underlying microscopic model. In statistical and high energy physics, the latter governs the time (inverse energy) dependence of  the interaction strength among fundamental particles. Here it regulates infectious interactions.  
 
Thus, the dictionary between the eRG equation for the epidemic strength $I(t)$ in an isolated region of the world  \cite{DellaMorte:2020wlc} and the high-energy physics analog is
 \begin{equation}
 \label{eq:beta0}
- \beta_{\rm eRG} (I(t)) =  \frac{d I(t)}{dt}  = {\gamma} \, I  \left( 1 - \frac{I}{A} \right)^{2p}\,, 
\end{equation}  
 whose solution, for $2p=1$, is a familiar logistic-like function
\begin{equation}
\frac{I (t)}{A} =  \frac{e^{{\gamma} t}}{b + e^{{\gamma} t}}\,.
\end{equation}
This solution can effectively describe one wave of the epidemic, and is related to the SIR model via time-dependent parameters \cite{DellaMorte:2020qry}.
The dynamics encoded in Eq.~\eqref{eq:beta0} is that of a system that flows from an Ultra-Violet fixed point at $t=-\infty$, where $I = 0$, to an Infra-Red one where $I = A$, with $A$ being the total number of infected cases at the end of the wave. 
The parameter $\gamma$ is an effective infection rate, characterising the diffusion slope. Both the equation and the solution show that $\gamma$ can be absorbed in a redefinition of time, $\tau = \gamma t$, so that $\gamma$ can also be interpreted as a \emph{time-dilation} factor, which differs by region and depends on external factors (social, demographic and relative to non-pharmaceutical interventions) which can slow down or accelerate the diffusion of the virus.
Finally, $b$ shifts the solution in time to match the beginning of the epidemic in each region, as shown by the equivalence $b \to 1$ with $\gamma t \to \gamma t - \ln b$. Further details, including what parameter influences the {\it flattening of the curve} and location of the inflection point and its properties can be found in \cite{mcguigan2020pandemic} and in \cite{DellaMorte:2020wlc}.

The presence of a truly interacting fixed point at large times predicts that the total number of cases approaches a constant value, while the new cases drop to zero. This feature can be associated to the system approaching a time scale-invariant state. Most regions affected by COVID-19, however, show a slightly different dynamics: in fact, the infection stabilises to an endemic period of constant growth, where the total cases continue growing linearly. This region can be associated to a breaking of the time scale-invariance \cite{cacciapaglia2020evidence}, which does not allow the system to approach the fixed point and generates the \emph{strolling} regime of linear growth. To account for this feature, the beta function in  \eqref{eq:beta0} can be extended as follows: 
\begin{equation}
- \beta_{\rm CeRG} (\iota) = \frac{d \iota}{d\tau} =  \iota  \left[  \left(1 - \iota \right)^2  - \delta\right]^p \, , 
\label{eq:beta1}
\end{equation}
where we have used normalised variable $\iota = I/A$ and $\tau = \gamma t$. The two new parameters $\delta$ and $p$ are real numbers, with $p$ positive. 
For negative $\delta$, the second factor has complex zeros, so that the system cannot approach a UV fixed point but, for small $|\delta|$, it lingers near the would-be fixed point $\iota \approx 1$ for a long time, inversely proportional to $|\delta|$.  This feature can explain the origin of the strolling \cite{cacciapaglia2020evidence}.
In this work, we will use the new Complex eRG (CeRG) approach to describe multi-wave dynamics.

The solution of Eq.~\eqref{eq:beta1} still encodes a single wave, followed by a strolling period. The latter signals an instability of the system, and precedes a new exponential increase of new cases.
The mathematical model can be extended to describe multiple waves by endowing the CeRG beta function with new (quasi) zeros as follows:
   \begin{equation}
- \beta_{\rm multiwaves} (\iota)  =   \iota \left[  \left(1 - {\iota} \right)^2  - \delta_0 \right]^{p_0} \; \prod_{\rho=1}^w \left[\left( 1-\zeta_\rho \iota \right)^2 - \delta_\rho \right]^{p_\rho}\ ,
 \label{eq:beta2wave}
\end{equation}
with $\zeta_\rho < 1$, $0 < -\delta_\rho\ll 1$ and $p_\rho >0$. We have that $w+1$ is the total number of pandemic waves, and that the diffusion will stop if $\delta_w = 0$, with a total number of infected cases equals $I (\infty) = A/\zeta_w$.     

For a case with two waves, $w = 1$, the value of $\iota$ at the two peaks (where the new infected cases reach a local maximum) can be determined by the zeros of the derivative of the beta function, $\partial \beta_{\rm multiwaves}/d\iota=0$. Setting $\delta_i = 0$ for simplicity (and because their value is very small and numerically irrelevant), we find
\beq
\iota_{\rm max}^{0/1} =\frac{1+\zeta_1 + 2 p_0 + 2 p_1 \zeta_1}{2 \zeta_1 (1+2 p_0 + 2 p_1)} \left( 1 \mp \sqrt{1-2 \frac{2 \zeta_1 (1+2 p_0 + 2 p_1)}{(1+\zeta_1 + 2 p_0 + 2 p_1 \zeta_1)^2}} \right)\,.
\eeq
From the definition of the beta-function, the time between the two peaks can be computed as follows:
\beq
\Delta \tau_{\rm peaks} = \int_{\iota_{\rm max}^0}^{\iota_{\rm max}^1} \frac{d x}{- \beta_{\rm multiwaves} (x)}\,.
\eeq
Note that the dependence on $\delta_1$ is negligible, because the domain of the integral is always far from the second set of complex zeros.
The value of $\delta$ is directly related to the number of new cases during the strolling, given by
\beq
S_{\rm t} = \left. \frac{d \iota}{d \tau} \right|_{\rm strolling} \approx - \beta_{\rm multiwaves} (1) = (- \delta_0)^p (1-\zeta_1)^{2 p_1}\,.
\eeq

\subsection*{Interacting regions}

Interactions among different regions of the world can be taken into account by adding to the beta function for each country $j$ the following interaction term \cite{Cacciapaglia:2020mjf}:
 \beq
\frac{\delta {I}_j (t)}{A_j\ \gamma_j \delta t} = \sum_l \frac{k_{jl}}{\gamma_j\ n_{mj}}  \frac{ {I}_l (t) - {I}_j (t)}{A_j}\,,
\label{eq:deltaI}
\eeq
where $n_{mj}$ is the population of region-$j$ in millions. The matrix $k_{jl}$ is proportional to the numbers of travellers between each of the regions considered in the interaction.

\subsection*{Extracting the parameters from the data}

We first fit the first and second wave (if reached) by use of independent solutions to the eRG framework. This exercise will allow us to compare the two waves, in particular the infection rates $\gamma_{\rm eRG}$, which are independent on the normalisation of the cumulative number of infected. 
We use open-source data from the online repository \href{https://ourworldindata.org/}{Ourworldindata.org}. 
We first identify the first wave, within the dates indicated in Table \ref{tab:2}, and fit the data by use of a simple logistic function:
\begin{equation}
    I(t) = \frac{A e^{\gamma_{\rm eRG} t}}{b + e^{\gamma_{\rm eRG} t}}\,.
\end{equation}
We then fit the data after the end date of the first wave (having subtracted the final number of cases) with a second independent logistic function plus a linearly growing term, which characterises the strolling period. If the second wave is absent, we simply fit the linear strolling. It is well known \cite{DellaMorte:2020wlc} that the fit is optimal only after the peak is achieved, so we tune the fit to reproduce the expected peak in some cases. In two cases we take into account spurious features in the data: for France, a sudden drop in the number of new cases can be observed after November 8, which is accountable by a decrease in the number of tests plus a reduced positivity rate; for Spain, the initial growth is due to a hotspot in Catalunya, followed by the spread to the rest of the country. Conservatively, we only fit the second higher sub-peak.

\begin{table}[tb!] \begin{center}
\begin{tabular}{|c|c|c|c|c|c|c|c|c|}
\hline
 & \multicolumn{8}{c|}{CeRG parameters}  \\ \hline
country  & $A$ & $\gamma$  &  $\delta_1$ &  $p_0$ & $p_1$ & $\zeta_1$ & 1st wave start & 1st wave end\\ 
\hline
France & $61 \cdot 10^3$ & $0.058$ & $1.0 \cdot 10^{-3}$ & $0.55$ & $0.6$ & $0.5$ & 2020-02-28 & 2020-05-08 \\
Italy & $41 \cdot 10^3$ & $0.091$ & $1.0 \cdot 10^{-3}$ & $0.55$ & $0.6$ & $0.5$ & 2020-02-18 & 2020-05-18\\
UK & $26 \cdot 10^3$ & $0.082$ & $1.0 \cdot 10^{-3}$ & $0.55$ & $0.6$ & $0.5$ & 2020-03-09 & 2020-06-17\\
Germany & $22 \cdot 10^3$ & $0.070$ & $1.0 \cdot 10^{-3}$ & $0.55$ & $0.6$ & $0.5$ & 2020-03-09 & 2020-05-18\\
Spain & $36 \cdot 10^3$ & $0.078$ & $1.0 \cdot 10^{-3}$ & $0.55$ & $0.6$ & $0.5$ & 2020-02-28 & 2020-05-18\\
Switzerland & $40 \cdot 10^3$ & $0.127$ & $1.0 \cdot 10^{-3}$ & $0.55$ & $0.6$ & $0.5$ & 2020-03-09 & 2020-04-28\\
Netherlands & $31 \cdot 10^3$ & $0.092$ & $1.0 \cdot 10^{-3}$ & $0.55$ & $0.6$ & $0.5$ & 2020-03-09 & 2020-05-18\\
Belgium & $45 \cdot 10^3$ & $0.157$ & $1.0 \cdot 10^{-3}$ & $0.55$ & $0.6$ & $0.5$ & 2020-03-09 & 2020-05-18\\
Denmark & $15 \cdot 10^3$ & $0.071$ & $1.0 \cdot 10^{-3}$ & $0.55$ & $0.6$ & $0.5$ & 2020-03-04 & 2020-05-28\\
Iceland & $11 \cdot 10^3$ & $0.114$ & $1.0 \cdot 10^{-3}$ & $0.55$ & $0.6$ & $0.5$ & 2020-03-04 & 2020-04-28\\
Canada  & $17 \cdot 10^3$ & $0.055$ & $1.0 \cdot 10^{-3}$ & $0.55$ & $0.6$ & $0.5$ & 2020-03-19 & 2020-06-17\\
\hline
South Africa & $12 \cdot 10^3$ & $0.102$ & $2.2 \cdot 10^{-3}$ & $0.56$ & $0.6$ & $0.45$ & 2020-04-16 & 2020-09-03\\
Bolivia & $13 \cdot 10^3$ & $0.057$ & $1.2 \cdot 10^{-4}$ & $0.51$ & $0.6$ & $0.45$ & 2020-04-20 & 2020-10-27\\
Saudi Arabia & $10 \cdot 10^3$ & $0.058$ & $1.3 \cdot 10^{-3}$ & $0.51$ & $0.6$ & $0.5$& 2020-03-21 & 2020-10-07\\
\hline
Australia & $772$ & $0.122$ & $8.4 \cdot 10^{-6}$ & $0.52$ & $0.6$ & $0.5$ & 2020-02-28 & 2020-04-23\\
Japan & $500$ & $0.103$ & $4.9 \cdot 10^{-2}$ & $0.7$ & $0.6$ & $0.4$ & 2020-02-28 & 2020-05-18\\
South Korea & $180$ & $0.191$ & $5.7 \cdot 10^{-3}$ & $0.6$ & $0.6$ & $0.45$ & 2020-02-18 & 2020-04-08\\ 
\hline
\end{tabular}
\caption{{\bf Parameters for the CeRG forecast for the third waves.}   \label{tab:2}}
\end{center}
\end{table}

As a second step, we tune the CeRG model, without interactions, to reproduce the last wave in order to forecast the arrival of the next one. The beginning of the wave (after subtracting any initial data) allows to determine $\gamma$, while $A_0$ and $p_0$ are determined by the end of the wave and the peak region. For countries where the second wave is still ongoing, we fix $p_0 = 0.55$, while its value is fitted in other cases. Furthermore, we fix the strolling parameter $S_t = 0.01$, or fit it to the data for countries that are currently in the strolling regime. The parameters obtained by this tuning are reported in Table \ref{tab:2}, and have been used to obtain the peak dates and in the plots.

\subsection*{Geographical uniformity indicator}

To investigate potential differences between waves due to the geographical spread of the virus, we define an uniformity indicator for each country by studying the number of new cases in various regions over a period of time around the peak.
The indicator is inspired to a $\chi^2$-variable, measuring how far is the actual distribution of cases compared to a perfectly uniform distribution, and it is defined as follows :
\begin{equation}
    \chi^2 (\Delta t) = \frac{1}{n_r}\sum^{n_r}_{i=1} \left(\frac{I_{ri}'(\Delta t)}{\left<I'_r (\Delta t)\right>} - 
     1\right)^2\,,
\end{equation}
where $I'_{ri} (\Delta t)$ is the number of new infected cases in the region $ri$ over a time interval $\Delta t$, and $\left<I'_r (\Delta t)\right>$ indicates the mean over the whole set of regions. We normalise the parameter to the number of regions in order to obtain comparable values. The smaller is $\chi^2$, the more uniform is the diffusion of the virus.
We also noted that the value of the uniformity indicator reaches local minima at the peak of each wave, thus the values reported for starting waves are larger than the final one.

To compute the values, we only used open-source data available on various online repositories. For Japan, we obtained the local data for the 46 prefectures from \href{https://toyokeizai.net/sp/visual/tko/covid19/en.html}{Toyokeizai.net}. In the analysis we omitted Okinawa because of its distance from the mainland. For Germany (16 Landen), Italy (20 regions), Spain (17 regions) and England (9 regions), we obtained data collected in periods of about 4 weeks from \href{https://www.citypopulation.de}{Citypopulation.de}. For France, as the local data on the new number of infected is not available before May 13, 2020, we used information for the number of hospitalisations in the 94 continental departments from \href{ https://www.data.gouv.fr/fr/datasets/donnees-hospitalieres-relatives-a-lepidemie-de-covid-19/}{Data.gouv.fr}. Finally, the data for each state of the US is obtained from \href{https://ourworldindata.org/}{Ourworldindata.org}.

For Japan and the US, the result for the third wave overestimates the one at the peak, which is not reached yet.

\section*{Additional plots}

\begin{figure}[tb!]
\begin{center}
\includegraphics[width=.99 \textwidth]{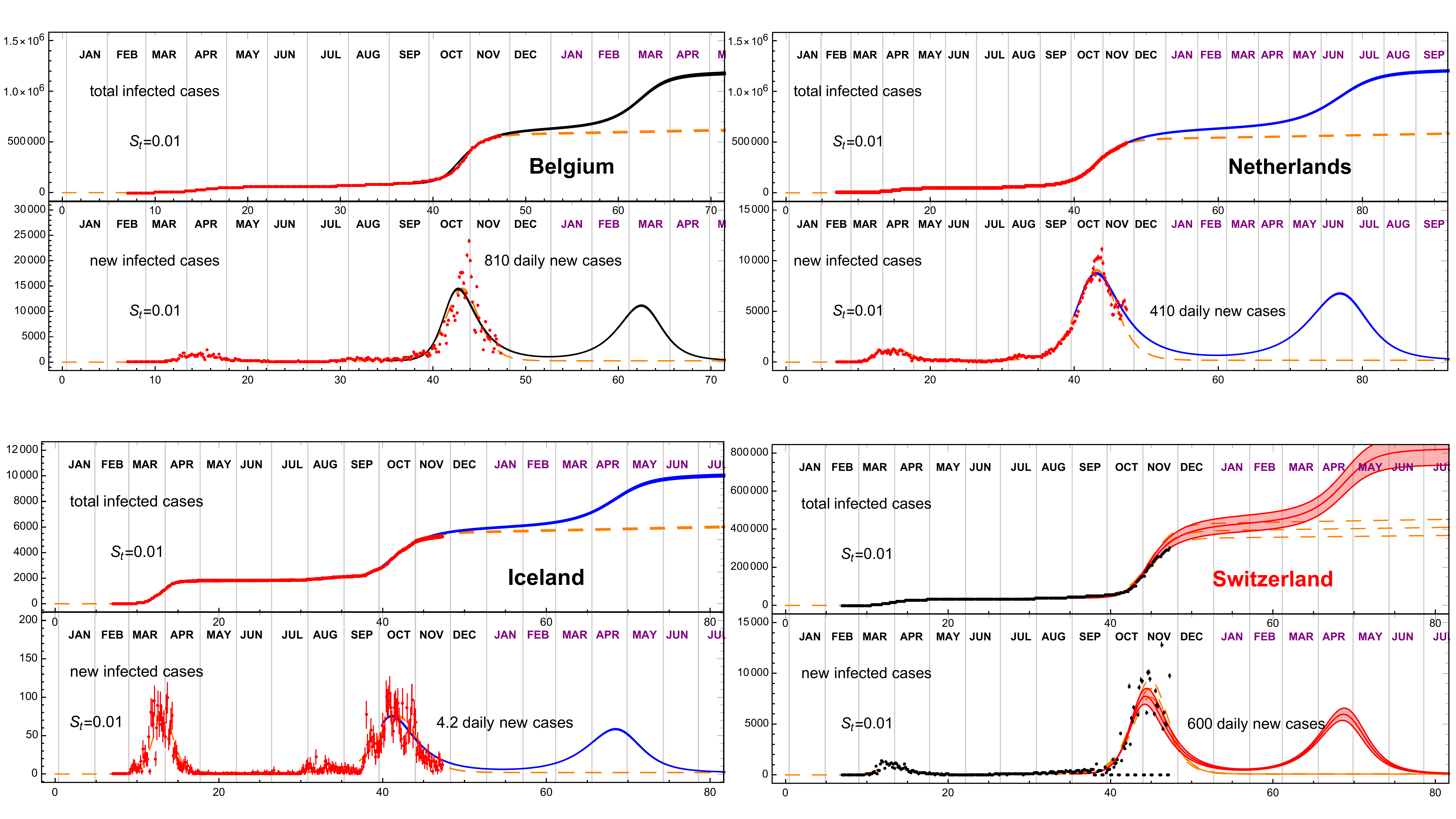}
\end{center}
\caption{{\bf Strolling as a precursor of a COVID-19 third wave.} Plots showing the CeRG model forecast for the future wave for a selection of European countries. The prediction corresponds to $S_{\rm t} = 0.01$, and is compared to the data (adjourned to November 23) and the eRG fit from Table \ref{tab:1} (dashed). }
\label{fig:4}
\end{figure}

\begin{figure}[tb!]
\begin{center}
\includegraphics[width=.99 \textwidth]{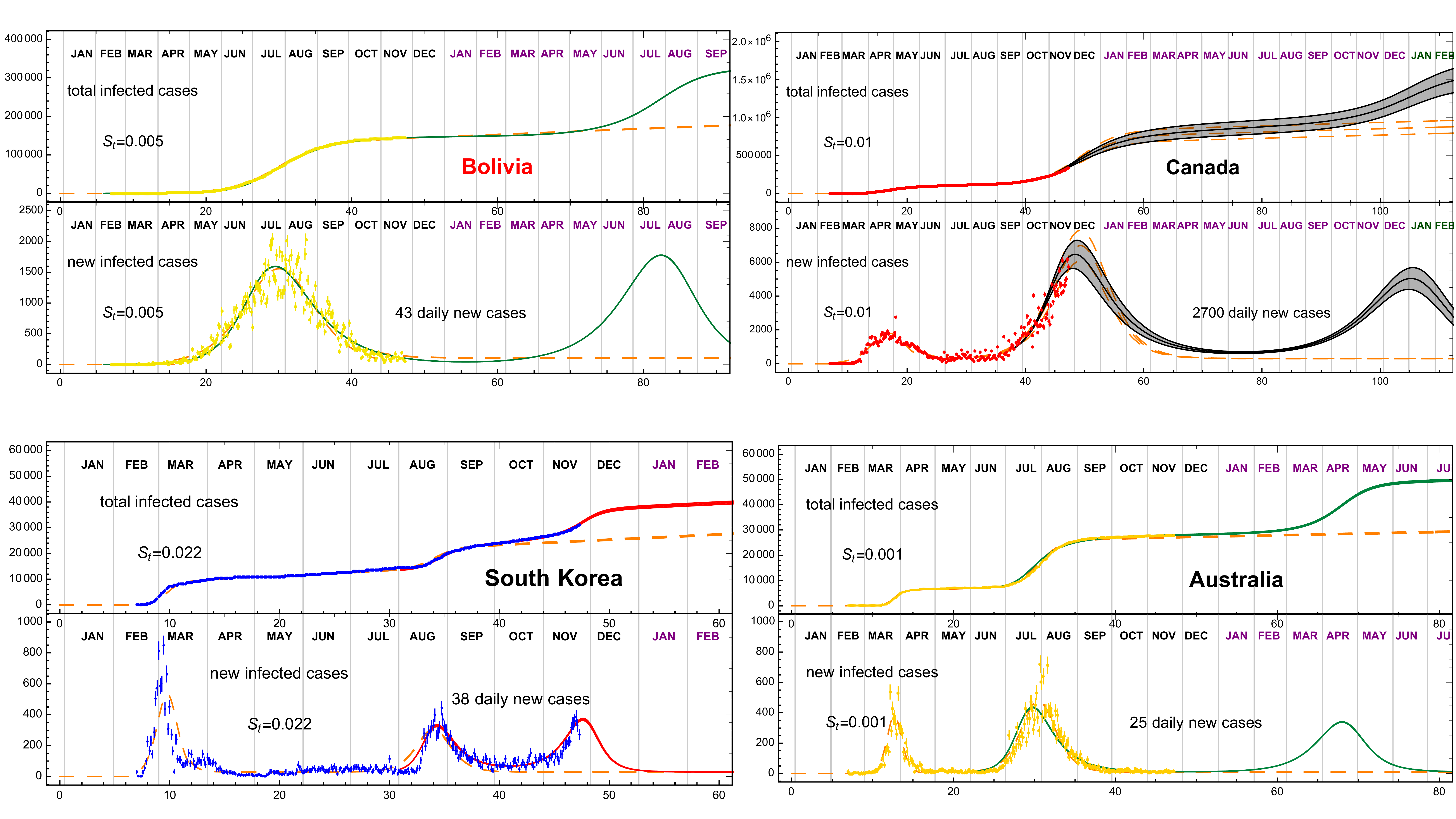}
\end{center}
\caption{{\bf Strolling as a precursor of a COVID-19 third wave.} Same as Fig. \ref{fig:4} for the remaining non-European countries. The prediction corresponds to $S_{\rm t} = 0.01$ for Canada, while the strolling parameter is fitted to the data for Bolivia, South Korea and Australia, and is compared to the data (adjourned to November 23) and the eRG fit from Table \ref{tab:1} (dashed). }
\label{fig:5}
\end{figure}

In Figs \ref{fig:4} and \ref{fig:5} we provide a visualisation of the CeRG forecast for the next wave for the countries not shown in the main text.

\bibliography{biblio}

\end{document}